\documentclass[aps, prl, superscriptaddress, reprint,aps,showpacs]{revtex4-1}

\usepackage{graphicx}
\usepackage{dcolumn}
\usepackage{bm}
\usepackage{color}

\newcommand{\VOtwo}[0]{VO$_2$ }
\newcommand{\votwo}[0]{VO$_2$}

\def\w {\omega}

\begin{document}

\title{Instantaneous bandgap collapse in photoexcited monoclinic VO$_2$ \\
due to photocarrier doping}

\author{Daniel Wegkamp}
\affiliation{Fritz-Haber-Institut der MPG, Faradayweg 4-6, 14195 Berlin, Germany}

\author{Marc Herzog}
\affiliation{Fritz-Haber-Institut der MPG, Faradayweg 4-6, 14195 Berlin, Germany}

\author{Lede Xian}
\affiliation{Nano-Bio Spectroscopy group, Universidad del Pa\'is Vasco CFM CSIC-UPV/EHU-MPC \& DIPC, 20018 San Sebasti\'an, Spain}
\affiliation{European Theoretical Spectroscopy Facility (ETSF)}

\author{Matteo Gatti}
\affiliation{Laboratoire des Solides Irradi\'es, \'Ecole Polytechnique, CNRS-CEA/DSM,  F-91128 Palaiseau, France}
\affiliation{European Theoretical Spectroscopy Facility (ETSF)}
\affiliation{Synchrotron SOLEIL, L'Orme des Merisiers, Saint-Aubin, BP 48, F-91192 Gif-sur-Yvette, France}

\author{Pierluigi Cudazzo}
\affiliation{Nano-Bio Spectroscopy group, Universidad del Pa\'is Vasco CFM CSIC-UPV/EHU-MPC \& DIPC, 20018 San Sebasti\'an, Spain}
\affiliation{European Theoretical Spectroscopy Facility (ETSF)}

\author{Christina L. McGahan}
\affiliation{Dept. of Physics and Astronomy and Interdisciplinary Materials Science Program, Vanderbilt University, TN 37235-1807, USA}

\author{Robert E. Marvel}
\affiliation{Dept. of Physics and Astronomy and Interdisciplinary Materials Science Program, Vanderbilt University, TN 37235-1807, USA}

\author{Richard F. Haglund, Jr.}
\affiliation{Dept. of Physics and Astronomy and Interdisciplinary Materials Science Program, Vanderbilt University, TN 37235-1807, USA}

\author{Angel Rubio}
\affiliation{Nano-Bio Spectroscopy group, Universidad del Pa\'is Vasco CFM CSIC-UPV/EHU-MPC \& DIPC, 20018 San Sebasti\'an, Spain}
\affiliation{Max Planck Institute for the Structure and Dynamics of Matter, Hamburg, Germany}
\affiliation{European Theoretical Spectroscopy Facility (ETSF)}
\affiliation{Fritz-Haber-Institut der MPG, Faradayweg 4-6, 14195 Berlin, Germany}

\author{Martin Wolf}
\affiliation{Fritz-Haber-Institut der MPG, Faradayweg 4-6, 14195 Berlin, Germany}

\author{Julia St\"{a}hler}
\email{staehler@fhi-berlin.mpg.de}
\affiliation{Fritz-Haber-Institut der MPG, Faradayweg 4-6, 14195 Berlin, Germany}

\date{\today}

\begin{abstract}
Using femtosecond time-resolved photoelectron spectroscopy we demonstrate that photoexcitation transforms monoclinic VO$_2$ quasi-instantaneously into a metal.
Thereby, we exclude an 80 femtosecond structural bottleneck for the photoinduced electronic phase transition of \votwo.
First-principles many-body perturbation theory calculations reveal a high sensitivity of the {\votwo} bandgap to variations of the dynamically screened Coulomb interaction, supporting a fully electronically driven isostructral insulator-to-metal transition.
We thus conclude that the ultrafast band structure renormalization is caused by photoexcitation of carriers from localized V $3d$ valence states, strongly changing the screening \emph{before} significant hot-carrier relaxation or ionic motion has occurred.
\end{abstract}


\maketitle

\begin{figure}
\includegraphics[]{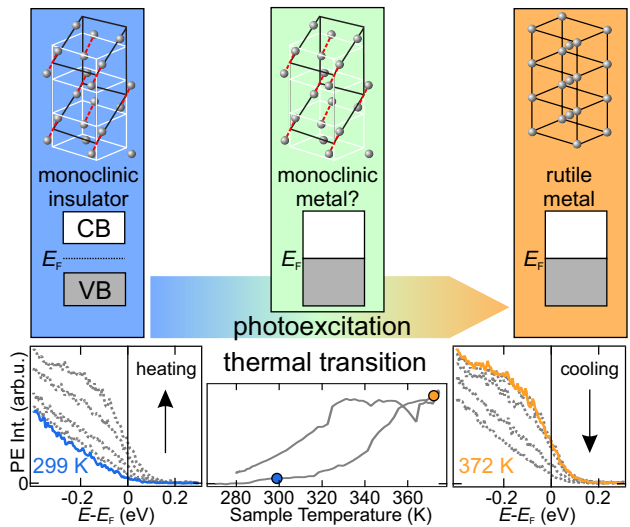}%
\caption{(color online) Top: The photoinduced PT from the insulating M$_1$ phase (left) to the metallic R phase (right) could proceed via a transient mM phase (green) or concurrently with the structural PT. Grey balls illustrate the V atom position. Bottom: Heating leads to the build-up of PE intensity in the bandgap (left) which is depleted upon cooling (right). Integrated PE intensity ($-0.2\text{eV} < E - E_{\text{F}} < 0.0\text{eV}$) exhibits a hysteresis (center). \label{fig1}}
\end{figure}

Since its discovery in 1959 \cite{Mori1959}, studies of the \VOtwo phase transition (PT) from a monoclinic (M$_1$) insulator (Fig.\ref{fig1}, top left) to a rutile (R) metal at $T_{\text{C}}=340K$ (Fig.\ref{fig1}, top right) have revolved around the central question \cite{Goodenough1971,Mott1975,Went1994,Biermann2005} of whether the \emph{crystallographic} PT is the major cause for the \emph{electronic} PT or if strong electron correlations are needed to explain the insulating low-$T$ phase. While the M$_1$ structure is a necessary condition for the insulating state below $T_{\text{C}}$, 
the existence of a \emph{monoclinic metal} (mM) and its relevance to the thermally driven PT is under current investigation \cite{Lim2006,Arcangeletti2007,Tao2012,Hegmann2012,Hsieh2014}. In particular, the role of carrier doping at temperatures close to $T_{\text{C}}$ by charge injection from the substrate or photoexcitation has been increasingly addressed \cite{Kim2004,Lim2006,Tao2012,Hada2012,Zhang2013,Siwick2014}.

One promising approach to disentangling the electronic and lattice contributions is to drive the PT non-thermally using ultrashort laser pulses in a pump-probe scheme. Time-resolved X-ray \cite{Cavalleri2001,Hada2010} and electron diffraction  \cite{Baum2007,Siwick2014} showed that the lattice structure reaches the R phase quasi-thermally after picoseconds to nanoseconds. Transient optical spectroscopies have probed photoinduced changes of the dielectric function in the THz \cite{Hilton2007,LeitenstorferPRL,LeitenstorferPRB}, near-IR \cite{Cavalleri2001,Wall2012,Hsieh2014,Hegmann2012} and visible range \cite{Wall2012}. The nonequilibrium state reached by photoexcitation (hereinafter \emph{transient phase}) differs from the two equilibrium phases, but eventually evolves to the R phase \cite{Cavalleri2001,Hada2010,Baum2007,Hilton2007,LeitenstorferPRL,LeitenstorferPRB,Wall2012,Cavalleri2004,vanVeen2013,Wall2013, Heinzmann2011,Shin2014}. The observation of a minimum rise time of $80\text{fs}$ in the optical response after strong excitation ($50\text{mJ/cm}^{2}$), described as a \emph{structural bottleneck} in \VOtwo \cite{Cavalleri2004}, challenged theory to describe the photoinduced crystallographic and electronic PT simultaneously \cite{vanVeen2013,Zhang2013}.

Time-resolved photoelectron spectroscopy (TR-PES) directly probes changes of the electronic structure. 
Previous PES studies of \VOtwo used high photon energies generating photoelectrons with large kinetic energies to study the dynamics of the electronic structure; however, with low repetition rate (50 Hz \cite{Heinzmann2011}) and inadequate time resolution ($>150\text{fs}$) the ultrafast dynamics of the electronic PT was inaccessible \cite{Shin2014}. Thus, fundamental questions - namely, how fast \emph{and why} the bandgap closes and whether this happens before or simultaneously with the crystallographic PT (Fig.\ref{fig1} top, center) - remained unanswered.

In this Letter, we use TR-PES to directly monitor the photoinduced changes of the density of states (DOS) around the Fermi energy $E_{\text{F}}$ which define the conduction properties of \votwo. We show that the insulating gap collapses during the exciting laser pulse ($<60\text{fs}$) with no sign of a structural bottleneck. The transient phase is an excited mM with carrier relaxation times on the order of $200\text{fs}$.
This interpretation of the experimental results is supported by first-principles calculations based on many-body perturbation theory.
They reveal that the bandgap in the M$_1$ phase is extremely sensitive to small changes in the occupation of the localized $d$ bands that alter the dynamically screened Coulomb interaction.
We thereby identify the origin of the metallization: photoexcitation induces holes by depletion of the V $3d$ orbital population \cite{Haverkort2005,Aetukuri2013} strongly affecting the screening and collapsing the bandgap.
Finally, the analysis demonstrates that, due to their strong localization, photoinduced holes are more effective than electrons in driving bandgap renormalization. In fact, hole doping can completely close the gap without the need of a structural change thus initiating a ``hole-driven insulator-to-metal transition''.

The 45 nm epitaxial \VOtwo film on a \emph{c}-cut sapphire crystal was grown at room temperature by pulsed laser ablation of a V target in an oxygen ambient \cite{Haglund2011}. For PES, it is kept under ultrahigh vacuum conditions and prepared by annealing cycles in an oxygen atmosphere. TR-PES is performed using a regeneratively amplified femtosecond laser working at a repetition rate of 40 kHz. A \emph{pump} pulse (h$\nu_{\text{pump}}= 1.54\text{eV}$) launches the non-equilibrium dynamics and its fourth harmonic (h$\nu_{\text{probe}}= 6.19\text{eV}$) serves as \emph{probe} pulse for photoemission. The incident pump fluence was 6.7(8) mJ/cm$^{2}$ (approx. 0.08 electrons per V atom) and probe fluence was kept below 8 $\mu$J/cm$^{2}$ to avoid charging of insulating \VOtwo and space-charge effects \cite{suppmat}.

\begin{figure}
\includegraphics[]{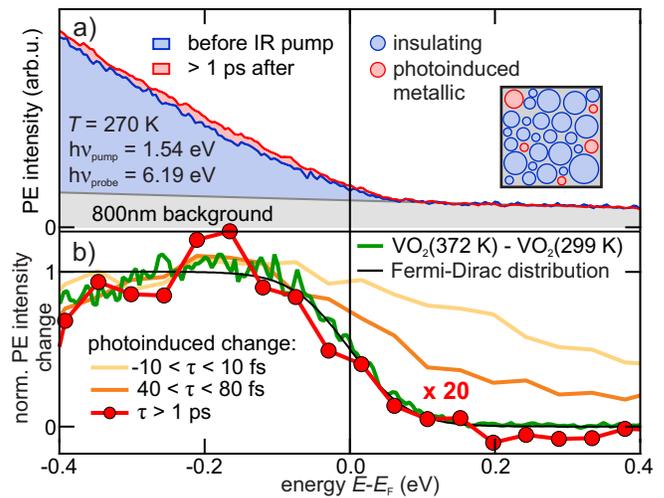}%
\caption{(color online) (a) PE spectra before (blue) and $>$1ps after pump (red, $6.7(8)\text{mJ/cm}^2$). Inset: Illustration of partial switching of selected domains. b) Comparison of thermally induced (green) and persistent photoinduced change (yellow, orange, red). The latter spectra were binned in energy at a step size of $\Delta E = 45\text{meV}$. \label{fig2}}
\end{figure}

Fig.\ref{fig1} bottom (left/right) depicts photoelectron (PE) spectra of \VOtwo in equilibrium at energies within the $\approx0.6$~eV bandgap of the insulating phase \cite{Koethe2006}.
The blue curve (299K) exhibits the high-energy tail of the \VOtwo valence band (VB). Heating the sample leads to a build-up of intensity (dotted curves, left) and a Fermi-Dirac (FD) like spectrum centered around $E_{\text{F}}$ at $372\text{K}$ (orange curve). This thermally induced PE intensity is suppressed upon cooling (dotted curves, right). The spectral weight below $E_{\text{F}}$ follows a hysteresis centered at $330\text{K}$ with a width of $25\text{K}$ (center panel) in line with optical experiments \cite{suppmat}. The difference between the high- and low-temperature PE spectra is plotted in Fig.\ref{fig2}b (green). Due to the excellent agreement of the thermally induced change of PE intensity with the \emph{parameter-free} FD distribution for $372\text{K}$ (black) \footnote{Broadened by the independently determined experimental energy resolution (90 meV).}, we conclude that, using $h\nu_{\text{probe}}= 6.19\text{eV}$, we are sensitive to the electronic PT in \votwo.

In order to elucidate how the photoinduced electronic PT evolves, we perform \emph{time-resolved} PES. We expect photoinduced changes on the order of $1-10\%$ of the thermally induced change, as reported for optical experiments at comparable pump fluences \cite{Cavalleri2001,Wall2013,Hsieh2014}. This is because only parts of the probed volume are transformed into the transient phase (Fig.\ref{fig2}a, inset) at excitation densities below the saturation regime ($F_{\text{sat}}\approx 4\cdot F_{TH}$, $F_{TH}$ threshold fluence for the PT) \cite{Cavalleri2004,Wall2013,Siwick2014} similar to the thermally driven PT \cite{Basov2007,Liu2013}. While optical experiments probe photoinduced changes due to carrier dynamics in the conduction band (CB) and VB of insulating \VOtwo \emph{and} close to $E_{\text{F}}$ of (potentially photoexcited) metallic \votwo, the energy selectivity of PES permits us to exclusively monitor photoinduced changes due to metallization by probing the dynamics \emph{in the gap} of insulating \votwo. Fig.\ref{fig2}a shows PE spectra before (blue) and $>1\text{ps}$ after the pump pulse. Indeed, the photoinduced change is very small. The difference is plotted in panel b (red markers) and compared to the thermally induced metallic spectral function (green).

As in optical experiments at comparable excitation fluences (see, e.g., Ref.\onlinecite{Wall2013}), the photoinduced signal is considerably smaller than the thermally induced one (here 5 \%). Yet, the curves show remarkable agreement, implying that the pump pulse has metallized individual grains of the sample (Fig.\ref{fig2}a, inset). Moreover, the signature of this transient metallic phase is practically identical with the thermally switched rutile \votwo. This direct observation of metallicity (defined by the presence of a Fermi edge) in photoexcited \VOtwo on ultrafast timescales goes beyond optical probes of metal-like dielectric functions, because those can be influenced strongly by highly excited electron-hole plasma in the CB and VB or structural changes that alter the dielectric function.

\begin{figure}
\includegraphics[]{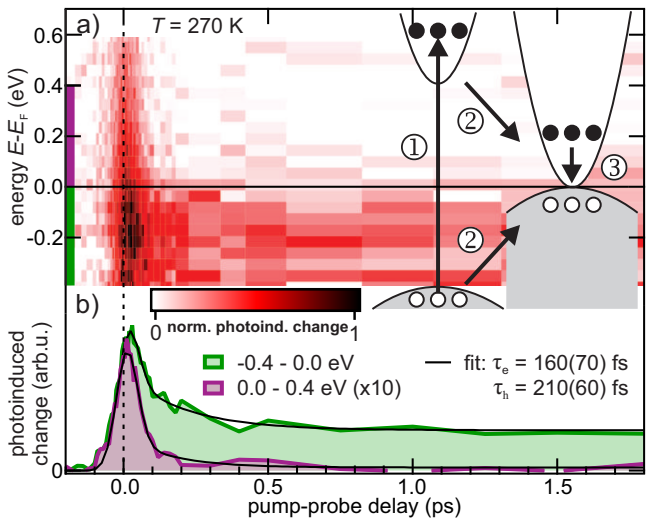}%
\caption{(color online) a) PE intensity change versus pump-probe delay close to $E_{\text{F}}$. Photoelectrons are detected immediately in the gap, showing the quasi-instantaneous collapse (cf. cartoon, energy axis is to scale). Integration of the PE intensity above (purple) and below (green) $E_{\text{F}}$ yields the respective transients in b). The empirical fit (black) quantifies the averaged hot electron (hole) lifetimes \cite{suppmat}.\label{fig3}}
\end{figure}

Fig.\ref{fig3} presents the ultrafast dynamics of the photoinduced electronic PT. The pump-induced change of PE intensity is depicted in false colors a) and is characterized by a fast component (fs timescales) and a long-lived intensity below $E_{\text{F}}$, which is spectrally equivalent to the photoinduced change in Fig.\ref{fig2}b. More precisely, the pump-induced intensity below $E_{\text{F}}$ represents the spectral signature of the transient metallic phase. Fully established at $1\text{ps}$, it is never modulated by coherent oscillations and remains unchanged for up to 400ps (not shown). This is noteworthy, as some time-resolved diffraction experiments on \VOtwo demonstrate an evolution of the atomic lattice over nanoseconds \cite{Cavalleri2001, Baum2007, Hada2010, Siwick2014}. The invariance of the TR-PE spectra on picosecond timescales shows that intermediate steps of the crystallographic PT have no effect on the FD distribution observed here, implying that the photoinduced electronic PT, i.e., the bandgap collapse, is completed before 1ps elapses.

Note that the photoinduced \emph{change} of the TR-PES signal in the gap below $E_{\text{F}}$ results \emph{only} from individual pump-induced metallized crystal domains, as photoexcited carrier dynamics in insulating \VOtwo would occur at different energies (in CB and VB). Thus, the PE intensity above $E_{\text{F}}$ (purple bar in Fig.\ref{fig3}a) corresponds to excited electrons in the transient (metallic) phase and PE intensity below $E_{\text{F}}$ (green bar) to dynamics in the occupied electronic band structure. The latter cannot result from defect states in the gap, as these would already be occupied in equilibrium.  The temporal evolution of integrated PE intensity in these energy windows is shown in Fig.\ref{fig3}b. Both traces are well fit by single exponential decays (bl\footnote{Photoinduced shifts of the chemical potential into VB (CB) are excluded, as they would be reflected in PE intensity at lower/higher energies than the sample holder $E_{\text{F}}$, respectively.}ack curves) with constant offsets \footnote{The fast dynamics on femtosecond timescales are overlapped with a lifetime-less two-photon photoemission (2PPE) signal of pump and probe pulses (through virtual states), which is accounted for in the fits by a delta function at time zero.}, convolved with the laser pulse envelope (duration: 61(5)fs \cite{suppmat}). The fits yield average decay constants $\tau_e=160(70)\text{fs}$ and $\tau_h=210(60)\text{fs}$. Importantly, the PE intensity in the gap is observed quasi-instantaneously at photoexcitation. We do \emph{not} observe a delayed rise of intensity below $E_{\text{F}}$ with a timescale of $80\text{fs}$ as expected for a structural bottleneck \cite{Cavalleri2004,suppmat}. On the contrary, the photoinduced PE intensity \emph{decreases} on a timescale of $210\text{fs}$ as expected for hole relaxation towards $E_{\text{F}}$ from lower energies. Fig.\ref{fig2}b depicts difference PE spectra at $t=0\text{fs}$ and $60\text{fs}$. They clearly display transiently occupied states close to $E_{\text{F}}$, superimposed by the lifetime-less intensity from two-photon absorption through virtual states. Therefore, the \emph{electronic} PT occurs with the photoexcitation of the \emph{electronic} structure and precedes any significant ionic motion towards the R phase.

In order to identify the physical mechanism for ultrafast metallization of photoexcited \votwo, we performed first-principles
calculations of the quasiparticle DOS within a many-body Green's-function approach \cite{suppmat,Onida2002}.
We adopted the $GW$ approximation for the self-energy $\Sigma$ \cite{Hedin1965}, because quasiparticle self-consistent $GW$, which naturally accounts for the localized character of the V $3d$ electrons, yields reliable quasiparticle band structures \cite{Gatti2007,*gattiPhD,Sakuma2008,Zhu2012} compared to experiments \cite{Koethe2006}. Alternatively, cluster DMFT (dynamical mean-field theory) is also able to describe the electronic structure of monoclinic \VOtwo \cite{Biermann2005,Laz2010}.
In the $GW$ approximation $\Sigma$ is given by the product of the one-particle Green's function $G$ and the dynamically screened Coulomb interaction $W(\w)=\epsilon^{-1}(\w)v$. Here $v$ is the bare Coulomb interaction and $\epsilon^{-1}$ is the inverse dielectric function calculated in the random-phase approximation including electron-hole and plasmon excitations \cite{suppmat}.

\begin{figure}
\includegraphics[width=\columnwidth]{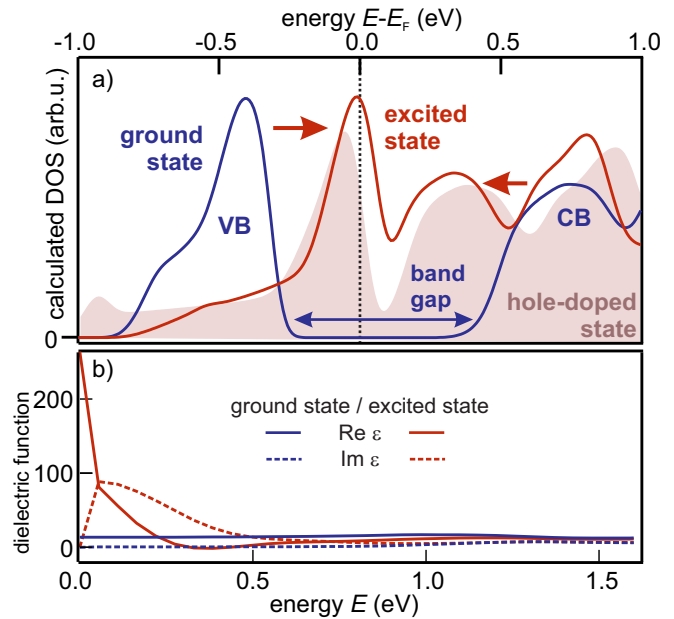}%
\caption{(color online) a) Calculated DOS (integrated over the entire Brillouin zone)
of the M$_1$ ground (blue) and excited states (red) broadened by the experimental resolution (90 meV).
b) Real and imaginary parts of the dielectric function for a representative small momentum transfer $\mathbf{q}=(1/6,0,0)$. \label{fig4}}
\end{figure}

The abruptness of the experimentally observed gap collapse justifies a Born-Oppenheimer approach with a ``frozen lattice''. In this spirit, we redistribute a portion of VB electrons equivalent to the experimental excitation density (0.075 electrons per V atom) to the unoccupied states. Note that our findings are robust with respect to excitation density \cite{suppmat}.
We then calculate the screened interaction ${\Delta}W$ that is changed by the presence of the additional carriers and the quasiparticle DOS
with the self-energy $\Delta\Sigma=G{\Delta}W$ \cite{suppmat,Oschlies1992,*Oschlies1995}. This redistribution of the electron and hole populations in the VB and CB is sufficient to lead to a collapse of the bandgap (Fig. \ref{fig4}a). In contrast to ordinary semiconductors where free-carrier doping leads to a moderate bandgap narrowing \cite{Abram1978,Wagner1985,Dou1997,Oschlies1992,*Oschlies1995,Spataru2004} and never results in a complete bandgap collapse purely electronically \cite{Spataru2004},
this extreme sensitivity of  \VOtwo to changes of the V $3d$ occupation is a distinctive and unique property of correlated materials \cite{Dagotto2005}.

The dynamical screening $\epsilon^{-1}(\w)$ in fact increases significantly in the low-energy region ($< 1.5$eV, Fig. \ref{fig4}b) due to creation of new VB-VB and CB-CB intraband electron-hole channels in the photoexcited system, while it remains almost unchanged in the high-energy region.
This finding is robust with respect to variations of the charge redistributions involving the depopulation of the V $3d$ bands \cite{suppmat}.
In order to unravel the microscopic mechanism at the origin of the bandgap collapse, we separately analyze the effect of changing the occupations of only VB or CB.
We find that hole doping at the top of VB alone indeed induces the bandgap breakdown \cite{suppmat} as suggested before \cite{Rini2008}.
We rationalize these findings by the fact that \VOtwo possesses an almost non-dispersive top VB corresponding to localized V $3d$ states \cite{Gatti2007,Eyert2002}.
Population changes of these states strongly enhance low-energy screening, leading to instantaneous metallization (bandgap closure). Note that pure electron doping also leads to a reduction of the bandgap but without metallization (no bandgap closure) \cite{suppmat}.

The relevant elementary processes are sketched in the inset of Fig.\ref{fig3}a. Absorption of pump photons lifts localized electrons from the top VB into the CB of insulating \votwo. This photocarrier doping causes an instantaneous breakdown of the gap \footnote{This is in agreement with Kim \textit{et al.} \cite{Lim2006,Kim2004}. However, in contrast to our work where we unambiguously prove the electronic origin of the PT, they use a weak laser to give the electronic PT a head start with respect to the structural PT when driving it thermally.} and excited electrons and holes subsequently relax at a slower rate towards equilibrium at $E_{\text{F}}$. The experimentally determined hot carrier relaxation times of order 200fs agree nicely with observations from other experiments. A similar time constant characterizes the incoherent time-dependent response of the conductivity in THz measurements of Pashkin \emph{et al.} \cite{LeitenstorferPRB}, which may well originate from excited carriers in the CB of \votwo. Also, pump-probe experiments of the \emph{transient phase} revealed that the optical response of photoexcited \VOtwo starts to resemble that of the thermally metallized sample after 200fs \cite{Wall2013}. Photoexcitation of \VOtwo creates an excited metal whose optical and electronic properties become similar to those of the thermally driven material only after the hot carriers have equilibrated. It is possible that the subsequent evolution of photoexcited \VOtwo towards the R phase occurs quasi-thermally, as the hot carriers thermalize with the lattice and heat it above $T_{\text{C}}$. Unlike the thermal PT, where lattice distortion and Coulomb interaction cooperatively drive the formation of the insulating gap, photoexcitation of the \emph{electronic} system instantaneously modifies the \emph{electronic correlations} causing the gap collapse, which is subsequently stabilized by the structural evolution.

In conclusion, the present experimental and theoretical work provides a comprehensive description of the elementary steps of the photoinduced electronic PT in \votwo. The bandgap of the insulating phase collapses instantaneously upon photoexcitation due to carrier doping,
revealing an ultrasensitivity of {\votwo} to variations of the screening by holes at the top of the V valence bands.
The bandgap collapse is followed by hot carrier relaxation in the transient metallic phase on a 200fs timescale and quasi-thermal evolution of the system towards the high-$T$ phase. The abrupt vanishing of the bandgap proves the absence of a structural bottleneck in the photoinduced electronic PT. Moreover, the electronic PT precedes the timescales observed for the crystallographic PT \cite{Hada2010,Baum2007} in excellent agreement with recent observations of a photoexcited mM-like \VOtwo \cite{Siwick2014}. These new insights into the character of the isostructural insulator-to-metal transition in \VOtwo provide not only a novel understanding of the physical mechanisms of the phenomenon, but also establish the basis for new experimental and theoretical studies of photoinduced dynamics in the transient metallic phase of \VOtwo as well as other correlated materials.

We gratefully acknowledge intense and fruitful discussions with S. Wall and A. Leitenstorfer and very useful comments from L. Perfetti about sample preparation. REM, CLM and RFH were supported by the National Science Foundation (DMR-1207507). AR, PC and LX acknowledge support by the European Research Council Advanced Grant DYNamo (ERC-2010-AdG-267374) and Grupos Consolidados UPV/EHU del Gobierno Vasco (IT-578-13). AR, PC, LX, MW and JS received support from the European Commission project CRONOS (Grant number 280879-2) and MG from a Marie Curie FP7 Integration Grant within the 7th European Union Framework Programme. Computational time was granted by GENCI (Project No. 544) and BSC ``Red Espanola de Supercomputacion''. DW acknowledges support from the Leibniz Graduate School DinL.

\bibliography{literature_05short}

\end{document}


\title{Supplementary materials}

\date{\today}
\maketitle

\section{Experimental details: Sample growth}

The \VOtwo film was deposited in an Epion PLD-3000 system using a Lambda Physik (Coherent COMPex 205) excimer laser (248~nm (KrF), 4~J/cm$^2$ per pulse, 25~Hz repetition rate, and nominal 25~ns pulse duration).  Prior to deposition, the chamber was pumped down to 9x10$^{-6}$~Torr.  A pure vanadium metal target was ablated at room temperature in an ultra-high purity oxygen ambient (1.1x10$^{-2}$~Torr, 2~sccm flow rate).  The laser beam was rastered across the rotating target while the substrate holder also rotated.  The average deposition rate was 0.3~\AA/s and the film thickness (nominal 45~nm) was verified using a Dektak profilometer.  The deposited film was annealed inside a tube furnace in 250~mTorr of O$_2$ at 723~K for 10 minutes.  After annealing, the film was allowed to cool before exposing them to ambient conditions. 

White-light transmission hysteresis measurements were taken to characterize the transmitted light as a function of temperature, controlled with a Peltier heater. Light from the near-blackbody tungsten lamp was focused onto the sample using a 5x (NA 0.20) microscope objective; the light transmitted through the film was collected with a 5x (NA 0.12) microscope objective and focused onto an InGaAs detector. The resulting optical hysteresis curve shows a switching contrast of 0.32, critical temperature of 326~K, and hysteresis width of 6~K, consistent with Beer's law for a stoichiometric and switching \VOtwo film and this spectral source.

\section{Experimental details: Photoelectron spectroscopy}

The experiments were performed at a repetition rate of the laser of 40~kHz, as the photoexcited \VOtwo did not fully recover upon arrival of the next pulse pair at higher repetition rates.  In contrast to optical experiments under ambient conditions, the reduced heat transport in UHV gave, even at the low repetition rate, an upper limit to the incident pump fluence of 7 mJ/cm$^{2}$. The \VOtwo sample was annealed in an oxygen atmosphere at 10$^{-4}$~mbar ($T = 600$~K, 30~min) until the photoemission (PE) spectra exposed the thermal phase transition (see Fig.1 in the main text). The emitted photoelectrons were detected using a hemispherical electron energy analyzer that was held at a fixed bias voltage of -0.5~eV with respect to the sample holder. All data shown are angle-integrated ($\pm7^\circ$). Photoelectron spectra are plotted as a function of energy with respect to the equilibrium Fermi level of the sample, which is in electrical contact with the sample holder. 

In contrast to traditional ARPES experiments using high photon energies, our experiments enable the clear distinction of sample charging and space charge effects that result from the cloud of photoemitted electrons: Using low photon energies, we always detect the complete PE spectrum starting at the low energy cut-off of secondary electrons with zero kinetic energy up to the electrons photoemitted from the Fermi energy $E_{\text{F}}$. While space charge effects generally lead to a broadening of the whole spectrum (fast electrons are accelerated, slow electrons become slower), charging of the sample leads to a shift of the entire spectrum to higher or lower energies due to negative (positive) charging. Both processes depend on the photon density of the \emph{probing} light pulse; the latter also depends on temperature (as it influences the electonic conductivity). To avoid such effects of the insulating \votwo, experiments were performed at sufficiently low probe photon flux (8~$\mu$J/cm$^{2}$) and sufficiently high temperatures (still $<T_{\text{C}}$) in order to maintain ample conductivity. 

\renewcommand\thefigure{S\arabic{figure}}
\begin{figure}
\includegraphics[]{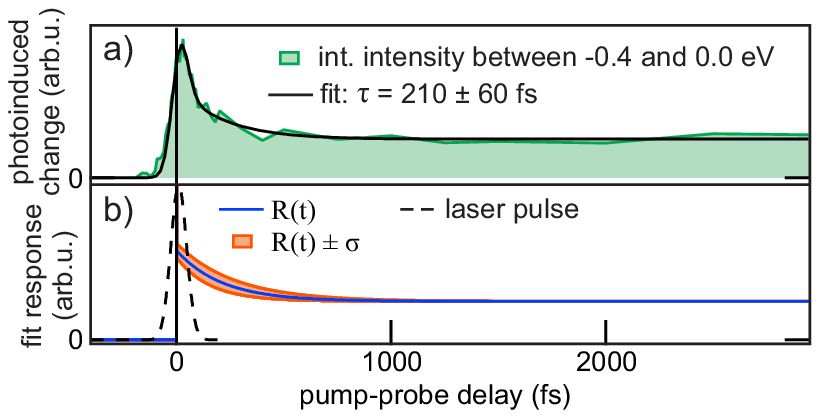}%
\caption{Population dynamics and fits \label{FigSexp}}
\end{figure}

\section{Fitting of time-resolved data}\label{sec:sec6}

The population dynamics (e.g. green curve in Fig.~\ref{FigSexp}a) are empirically fitted with the function 

\begin{equation}
R_{1}(t)=\Theta(t)\cdot(a+b\cdot e^{-t/\tau})+c\cdot\delta(t)
\end{equation}

where the first term is a single exponential decay with a constant offset starting at time zero (blue curve in Fig.~\ref{FigSexp}b) and the second term accounts for photoelectrons emitted after absorption of two photons via (virtual) states without lifetime. Convolution with the pump and probe laser pulses' envelopes (dashed curve) yields the black fit in Fig.~\ref{FigSexp}a. Their cross correlation (two Gaussians) is determined at high energies, and has a width of 89 fs, leading to a mean pulse duration of pump and probe pulses of 63 fs, which gives a conservative upper limit for the error of the decay time determination. Fig.~\ref{FigSexp} illustrates the deviation that results from a 60 fs error with the other fit parameters fixed (red).

\begin{figure}
\includegraphics[]{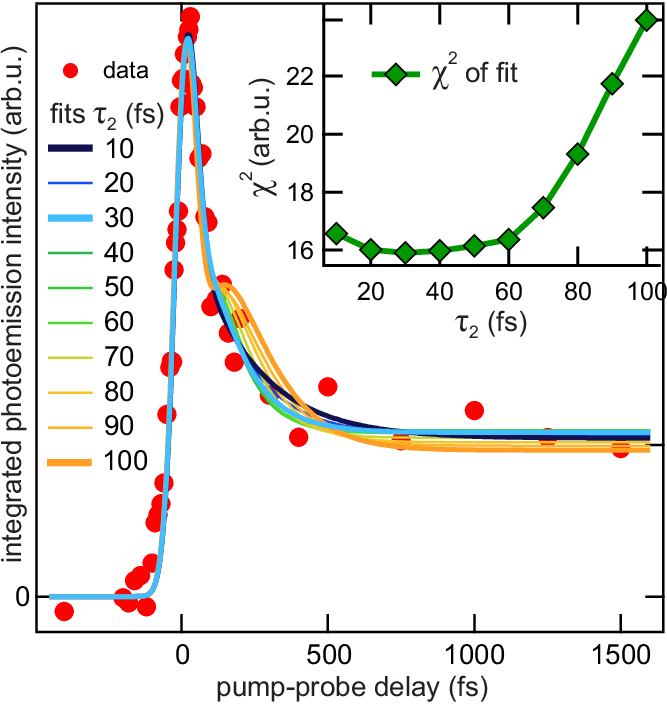}%
\caption{Fitting of the rise time \label{FigSexp2}}
\end{figure}

In order to test, whether the data is consistent with the 80 fs structural bottleneck, we also fitted the empirical function

\begin{equation}
R_{2}(t) = \theta(t)\left(A_1 e^{-t/\tau_1} - A_2 e^{-t/\tau_2} + C\right) + D\cdot \delta(t)
\label{eq:fitbelowfermi}
\end{equation} 

to the time-dependent change of PE intensity below $E_{\text{F}}$. The results are shown in Fig.~\ref{FigSexp2}. The first term empirically describes the population dynamics below $E_{\text{F}}$ while the second one accounts for the two-photon PE spike at $t=0$. Again, this function was convolved with the laser pulses' envelope as discussed above. Assuming that the DOS is zero before the pump pulse arrives, the population response must be zero at $t=0$ and then rise ($\tau_2$) as the structural bottleneck is passed. The decay ($\tau_1$) accounts for possible hot hole relaxation towards $E_{\text{F}}$. Thus, $A_1-A_2+C=0$ is a boundary condition for this fit. Fig.~\ref{FigSexp2} shows a series of different fits with fixed rise times $\tau_2$ and the corresponding $\chi^2$ in the inset. Obviously, the fits yield satisfactory results \emph{only} for rise times \emph{below} 60 fs, i.e. below the structural bottleneck time of 80 fs that was observed in optical experiments \cite{Cavalleri2004}.

\section{Theoretical framework}

In the present work we use the $GW$ self-energy approximation on top of a density-functional theory (DFT) calculation of VO$_2$ to account for correlation effects. This approach is known to provide an accurate description of both insulating and metallic phase of VO$_2$ \cite{Gatti2007}. In the $GW$ approximation (GWA) \cite{Hedin1965}  the self-energy $\Sigma$ is given by the convolution in frequency space:
\beq
\Sigma(\bfr_1,\bfr_2,\w) = \frac{i}{2\pi} \int d\w' e^{i\eta\w'} G(\bfr_1,\bfr_2,\w+\w')W(\bfr_1,\bfr_2,\w')
\label{sigma}
\eeq
between the time-ordered one-particle Green's function $G$:
\begin{equation} \label{G}
G(\br_1,\br_2,\omega)
 = \sum_i \frac{\phi_i(\br_1)\phi^*_i(\br_2)}{\omega - \epsilon_i + i \eta\sgn(\epsilon_i-\mu)},
\end{equation}
(where $\mu$ is the Fermi energy and $\eta \rightarrow 0^+$ )
and the dynamically screened Coulomb interaction $W$:
\begin{equation}\label{W}
W(\br_1,\br_2,\omega) = \int d\br_3 \, \epsilon^{-1}(\br_1,\br_3,\omega) v(\br_3-\br_2).
\end{equation}
Here $v$ is the (static) bare Coulomb interaction and $\epsilon^{-1}$ is the inverse dielectric function that describes the screening of $v$ through electron-hole and collective plasmon excitations.
In the GWA $\epsilon^{-1}$ is calculated in the random-phase approximation (RPA) from the independent-particle polarizability $\chi_0$:
\beq\label{epsilon}
\epsilon^{-1}(\w) = 1 + v \frac{\chi_0(\w)}{1 - v\chi_0(\w)}.
\eeq
(Here integrations/inversions over spatial coordinates are implicitly understood.)
Finally, the (time-ordered) independent-particle polarizability $\chi_0$ is:
\beq\label{chi0}
\chi_{0}(\bfr_1,\bfr_2,\w) =
\sum_{i,j}(f_{i}-f_{j})
\frac{\phi_i(\bfr_1)\phi_j^*(\bfr_1)\phi_i^*(\bfr_2)\phi_j(\bfr_2) }{\w-(\varepsilon_{i}-\varepsilon_{j})+i\eta\sgn(\varepsilon_{i}-\varepsilon_{j})},
\eeq
where $f_i$ are the occupation numbers.

In the standard approach, $G$ and $W$ are constructed using the Kohn-Sham eigenvalues and wavefunctions obtained in a DFT calculation with local density approximation (LDA). However, such calculations fail to reproduce the insulating nature of the monoclinic VO$_2$ (M$_1$ phase) \cite{Gatti2007}. Then we need to rely on more sophisticated orbital-dependent potentials to capture the correct electronic structure of both insulating and metallic phase. Thus in this work, the quasiparticle (QP) wavefunctions $\phi_i$  entering Eqs.~\eqref{G}-\eqref{chi0} are obtained from a QP self-consistent COHSEX calculation (see Refs. \onlinecite{Bruneval2006,Gatti2007}).
COHSEX is a static approximation to the GWA self-energy (Eq.~\eqref{sigma}) that is given by the sum of a Coulomb-hole (COH) and a screened-exchange (SEX) terms \cite{Hedin1965}.
The QP energies $\varepsilon_i$ appearing in Eq.~\eqref{G} are then calculated self-consistently
by solving the QP equation
\begin{eqnarray}
\left(-\frac{\nabla^2}{2}+V_{ext}(\br_1)+ V_{H}(\br_1)\right)\phi_i(\br_1)   \nonumber \\
+ \int d\br_2 \Sigma(\br_1,\br_2,\varepsilon_i)\phi_i(\br_2)  = \varepsilon_i\phi_i(\br_1),
\label{QP}
\end{eqnarray}
following a $GW_0$ scheme (i.e. $\phi_i$ and $W$ are kept fixed at the COHSEX level).
In Eq.~\eqref{QP} $V_{H}$ is the Hartree potential and $V_{ext}$ is the electron-ion interaction.
At self-consistency, the QP energies $\varepsilon_i$ determine the QP density of states (DOS), integrated over the whole Brillouin zone, that is plotted in the following figures (also the one-shot $G_0W_0$ results on top of COHSEX do not differ qualitatively from the $GW_0$ DOS). 

\section{Numerical details}

Here, we have used the Abinit code to perform the calculations \cite{abinit}. We have adopted the experimental crystal lattice of the monoclinic $M_1$ structure \cite{Longo1970} (atomic positions are kept frozen) and used Troullier-Martins pseudopotentials \cite{Martin1991} (with V $3s$ and $3p$ explicitly treated as valence electrons). A plane-wave basis set with cutoff of 180 Hartree has been employed. COHSEX results are obtained from Ref.~\onlinecite{Gatti2007}. Convergence has been achieved using 200 bands for the calculation of the screening, and 150 bands in the calculation of self-energy corrections. We have used 5007 plane waves to expand the wavefunctions entering the screening and 14999 plane waves for the wavefunctions entering the self-energy. We have calculated the frequency convolution in Eq.~\eqref{sigma} using an accurate contour-deformation technique with 60 frequencies along the real axis up to 1.5 Hartree and 10 frequencies along the imaginary axis. In the main part of the present article, the DOS are calculated with a 6$\times$6$\times$6 $\Gamma$-centered grid of $\bfk$ points. The results discussed in the following of this Supplementary Material are instead based on less expensive calculations performed with a 4$\times$4$\times$4 grid of $\bfk$ points that provides the same qualitative results as the more converged calculations with the 6$\times$6$\times$6 $\bfk$-grid.

In the time-resolved photoemission (TR-PE) experiment, the laser pump excites 0.08 electrons per V atom (i.e. 0.32 electrons per monoclinic unit cell), partially depleting V $3d$ valence states and partially occupying V $3d$ conduction states above the gap (the O $2p$ states instead cannot be excited for energy-conservation reasons: the top of the O $2p$ band has a binding energy that is larger than the energy of the laser pump). In order to simulate pump-probe TR-PE experiments, formally one should solve the non-equilibrium Baym-Kadanoff (BK) equations \cite{Stefanucci2013}. 
However, for the goals of the present work we investigate the effect of the photoinduced instantaneous modification of the occupations $f_i$ on the  measured QP DOS through the change of the screened Coulomb interaction ${\Delta}W$, which is one of the ingredients in the solution of the BK equations and it provides a physical understanding of the microscopic process leading to the bandgap collapse discussed in the main text.
In fact, from Eq.~\eqref{chi0} we see that variations of the $f_i$ in the photoexcited states directly affect $\chi_0$ and hence the screening $\epsilon^{-1}$ through  Eq.~\eqref{epsilon}.
The QP DOS are recalculated using the GWA with a ${\Delta}W$  modified according to different occupation distributions $f_i$, i.e. with $\Sigma=G{\Delta}W$ \footnote{the study of the effects due to variations of $f_i$  on the Hartree potential ${\Delta}V_H$, for the change of the electron density, and on the Green's function ${\Delta}G$, for the change of the pole structure in Eq.~\eqref{G}, is left for a future work. We do not expect the conclusions of the present analysis will change once those contributions are included. }.
The differences between the resulting DOS are entirely due to different screened Coulomb interactions ${\Delta}W$ originating from different occupation distributions $f_i$.

\section{Extended analysis of the theoretical calculations}

\begin{figure*}[htbp]
\includegraphics[width=0.8\textwidth]{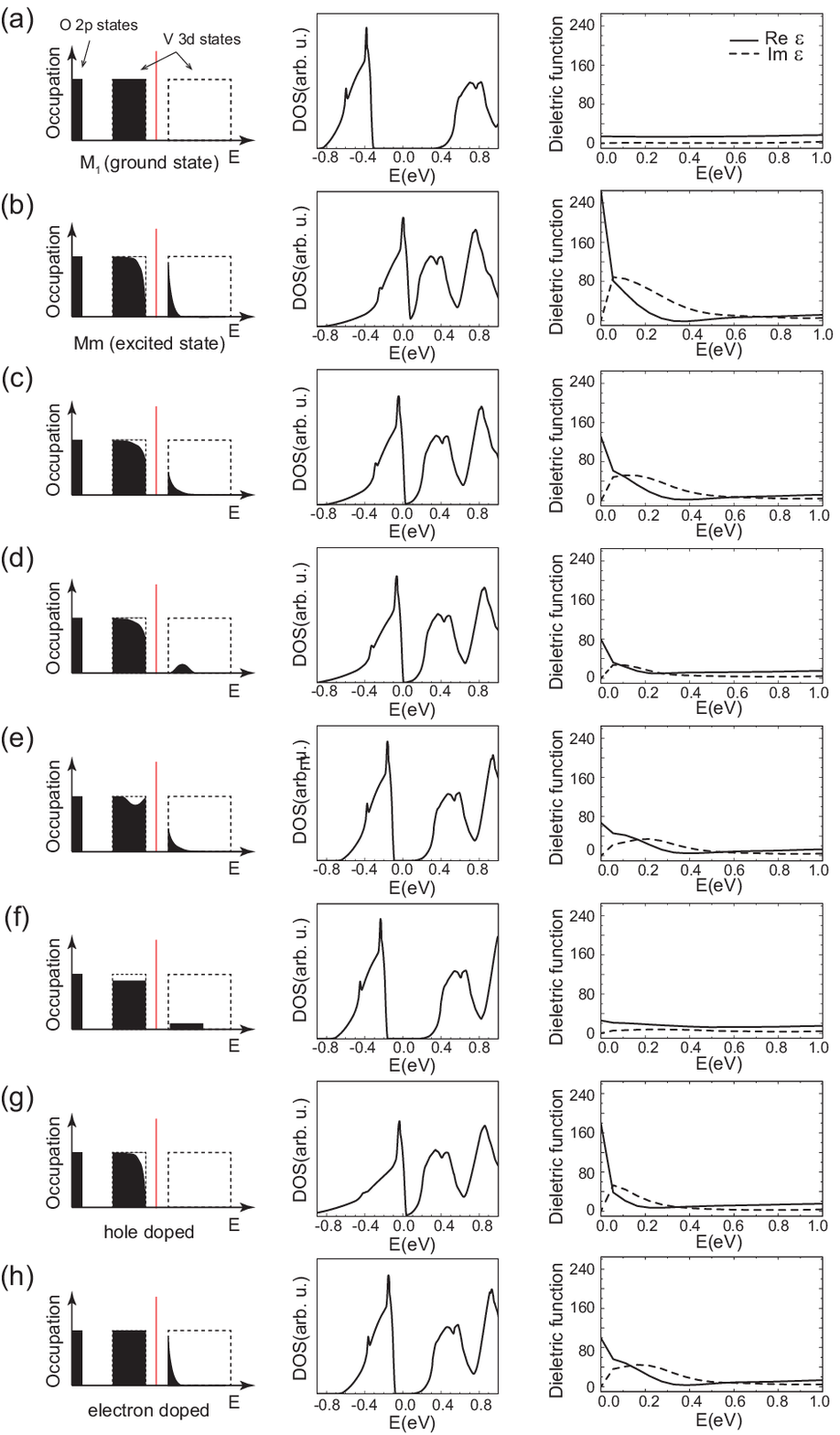}
\caption{(Left) Occupation distribution used to calculate the screened Coulomb interaction.
(Center) Quasiparticle density of states obtained from the given occupation distribution.
 (Right) Real and imaginary parts of the dielectric function calculated in RPA at $\bfq=(1/6,0,0)$.
 The ground state is represented in panel (a), photoinduced excited states in panels (b)-(f), hole and electron doping in panels (g) and (h) respectively. The calculations correspond to an electron/hole density of 0.075 carriers per V atom.}
\label{figs1}
\end{figure*}

We have calculated the QP DOS with some representative occupation distributions to analyze in detail the effect of the photoexcitation on the screening (neglecting electron-hole interactions). The results are shown in Fig.~\ref{figs1}. In particular, Fig.~\ref{figs1}(a, center) shows the QP DOS for the monoclinic $M_1$ phase in the ground state (i.e. before the photoexcitation) calculated in the GWA, reproducing previous results\cite{Gatti2007}.
It displays a gap of $\sim 0.6$ eV between the top-valence and bottom-conduction V $3d$ states, in agreement with earlier PE spectra \cite{Koethe2006}. In the ground state all the valence states below the Fermi level, comprising V $3d$ states and O $2p$ at lower energy, are fully occupied ($f_i=2$) and those above the gap are fully empty ($f_i=0$), see Fig.~\ref{figs1}(a, left).

Fig.~\ref{figs1}(b)-(f)(center) show the QP DOS obtained for different excited occupation distributions [see Fig.~\ref{figs1}(b)-(f)(left)], in all of which 0.075 electrons per V atom are excited from valence to conduction states.
 In Fig.~\ref{figs1}(b)-(c) the occupation distributions have the same  valence and conduction quasi-Fermi levels, but different effective electronic temperatures. In Fig.~\ref{figs1}(d)-(f) the distributions are non-thermal.

 In all cases with excited occupation distributions,  we find important changes in the QP DOS with respect to the ground state: there is always at least a  conspicuous bandgap narrowing. In Fig. S\ref{figs1}(b)-(d) we even observe a complete bandgap collapse.
 This finding is an evidence of the very high sensitivity of the {\votwo} band structure with respect to changes in the V $3d$ occupations.

 In order to understand the microscopic mechanism that leads to the bandgap collapse, in Fig.~\ref{figs1}(a)-(f)(right) we have also plotted the real and imaginary parts of the dielectric function $\epsilon(\bfq,\w)$, calculated in RPA in the low-energy range ($\w < 1.0$~eV) for a small momentum transfer $\bfq=(1/6,0,0)$ (for other $\bfq$ similar considerations apply).
 In the ground state, see Fig.~\ref{figs1}(a), both $\Re \epsilon$ and $\Im \epsilon$ are flat. In fact, in RPA there cannot be electron-hole transitions at energies below the fundamental direct bandgap ($\sim$ 0.8 eV).
 The change in the occupations,  see Fig.~\ref{figs1}(b)-(f), develops new peaks in $\Im \epsilon$ for $\w < 1.0$ eV that are linked to the opening of new electron-hole excitation channels at energies smaller than the ground-state bandgap ($\Re \epsilon$ is analogously modified as it is connected to $\Im \epsilon$ by Kramers-Kronig relations). This can be immediately realised from Eq.~\eqref{chi0}: the independent-particle polarizability $\chi_0$ have new poles for transitions between pairs of states for which $f_i \neq f_j$.

\begin{figure*}[htbp]
\includegraphics[width=0.6\textwidth]{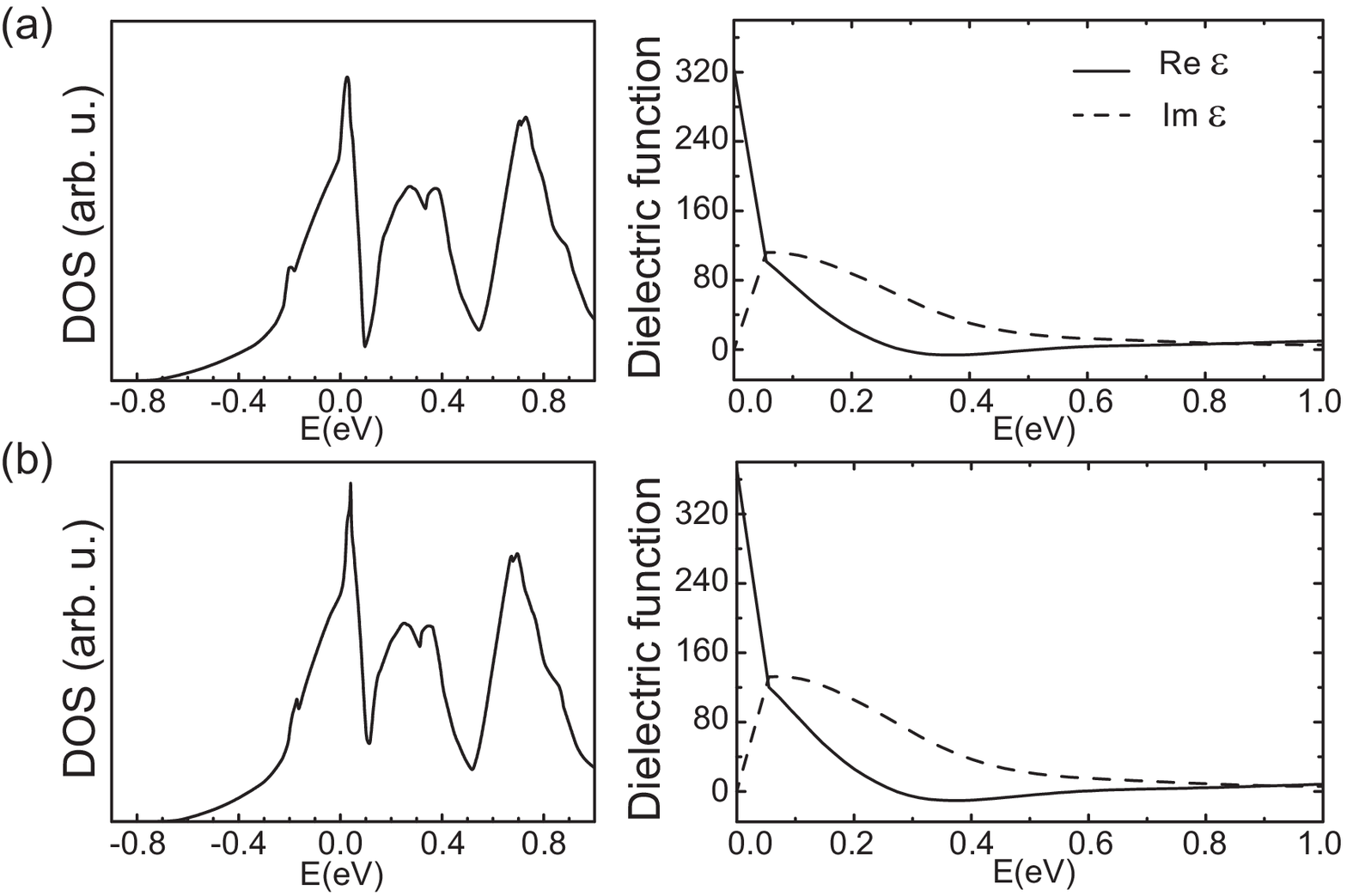}
\caption{Same as Fig.~\ref{figs1}(b), here with a larger number of excited electrons: (a) 0.1 and (b) 0.125 per V atom.}
\label{figs3}
\end{figure*}

 The new peaks in $\Im \epsilon$  are mainly due to transitions within the valence band (VB) and within the conduction band (CB) (``intraband transitions'').
 In fact, when the new occupation distribution remains flat, see Fig.~\ref{figs1}(f)(left), smaller structures are present in $\Im \epsilon$, see Fig.~\ref{figs1}(f)(right), which are only due to transitions from partially filled CB to completely empty CB. In turn this leads to a smaller bandgap narrowing. This holds to a certain extent also for Fig.~\ref{figs1}(e), where holes are created in the middle and not at the top of the VB. On the contrary, whenever intraband transitions are made possible, the screening is largely modified, leading to the bandgap collapse [independently of the electronic temperature, compare Figs.~S\ref{figs1}(b)-(c)].
 This finding is confirmed also when the excited electronic charge is larger, for example 0.1 and 0.125 electrons per V atom, see Fig.~\ref{figs3}.

The overall picture that emerges from this analysis is clear: the change of the occupations allows intraband transitions to modify the low-energy screening of the Coulomb interaction. In turn, the modified ${\Delta}W$ leads to a dramatic change of the electronic structure, revealing a high sensitivity of the electronic properties of {\votwo} to external changes.

\begin{figure*}[htbp]
\includegraphics[width=0.608\textwidth]{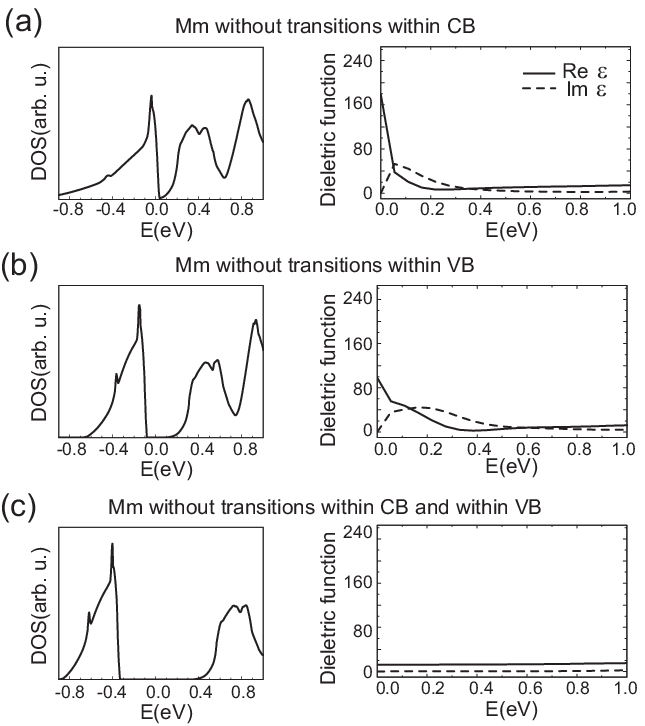}
\caption{Starting from the occupation distribution of Fig.~\ref{figs1}(b), intraband transitions have been suppressed (a) within the CB, (b) within the VB, and (c) in both the VB and the CB.}
\label{figs2}
\end{figure*}

In order to analyze the effect of intraband transitions within the VB  and the CB separately, we have followed two paths: (i) select the occupation distribution of Fig.~\ref{figs1}(b) and perform calculations where we remove intraband transitions within the CB or the VB from $\chi_0$ in Eq.~\eqref{chi0}; (ii) perform calculations with hole doping or electron doping with 0.075 electrons per V atom removed from the top VB or added to the bottom CB. These two approaches differ slightly by some interband transitions. The results of approach (i) are shown in Fig.~\ref{figs2}(a)-(c).
When all the  intraband transitions are suppressed, see Fig.~\ref{figs2}(c),  the peak in $\epsilon$ in the low-energy regions disappears. In this case, the bandgap remains open and is almost the same as in the ground state [see Fig.~\ref{figs1}(a)].
 When intraband transitions are allowed within the CB only, see Fig.~\ref{figs2}(b), the bandgap is reduced by 0.46 eV but not closed.
Finally, when intraband transitions are allowed within the VB only, see Fig.~\ref{figs2}(a), the bandgap collapses, nearly retrieving the result of the total calculation, see Fig.~\ref{figs1}(b).

Therefore, this analysis clearly shows that intraband transitions within VB play a crucial role in changing the screening and closing the bandgap. This is due to the fact that the top VB states are largely non-dispersive owing to their effective electronically one-dimensional character which is linked to the bonding states of the V dimers along the $c$-axis \cite{Eyert2002}. This peculiar property gives rise to an intense peak in the low-energy region of $\Im \epsilon$ when intraband channels are activated by the electron excitation.

While the intraband transitions are artificially suppressed in approach (i), the realistic means to realize those conditions in experiments are through electron or hole doping, which are simulated in approach (ii). The results of approach (ii) for an electron/hole density of 0.075 carriers per V atom are shown in Fig.~\ref{figs1}(g)-(h). In accordance with approach (i), we find that while for the pure electron doping the bandgap is strongly reduced but not completely closed, the change in the dielectric function due to hole doping is sufficient to close the bandgap. The same result holds for the higher electron/hole densities discussed in Fig.~\ref{figs3}, namely hole doping closes the bandgap but electron doping leads to a positive gap of about 0.1 eV. These findings thus suggest that, in addition to the photoinduced phase transition reported in this work, alternative experimental approaches employing pure hole doping also provoke metallization of {\votwo}. 

\bibliography{literature_SI}